\documentclass[12pt]{article}
\usepackage{amsmath,amssymb}

\def\today{\ifcase\month\or
  January\or February\or March\or April\or May\or June\or
  July\or August\or September\or October\or November\or December\fi
        \space \number\year}
\def\versionno{M.~Postma}
\catcode`\@=11
{\count255=\time\divide\count255 by 60
\xdef\hourmin{\number\count255}
\multiply\count255 by-60\advance\count255 by\time
\xdef\hourmin{\hourmin:\ifnum\count255<10 0\fi\the\count255}}
\def\ps@draft{\let\@mkboth\@gobbletwo
    \def\@oddhead{}
    \def\@oddfoot{\hbox to 7 cm{\tiny \versionno
       \hfil}\hskip -7cm\hfil\rm\thepage \hfil {\tiny\draftdate}}
    \def\@evenhead{}\let\@evenfoot\@oddfoot}
\def\draftdate{\number\day.\number\month.\number\year\ \ \ \hourmin}

\catcode`\@=12
\newcommand{\thisfile}[1]{\renewcommand{\versionno}[0]%
{M.~Postma --- File: #1.tex}}

\input epsf
\def\be{\begin{equation}}
\def\ee{\end{equation}}
\def\bea{\begin{eqnarray}}
\def\eea{\end{eqnarray}}

\renewcommand\({\left(}
\renewcommand\){\right)}
\renewcommand\[{\left[}

\newcommand\GeV{\,\mbox{GeV}}

\newcommand\mpl{M_{\rm pl}}

\begin{document}

\thisfile{ok}

\vskip 1cm
\begin{center}
{\Large \bf Inhomogeneous reheating scenario with low scale inflation
and/or MSSM flat directions
\vskip 0.2cm}
\vspace{0.4in}
{Marieke Postma}
\vspace{0.3in}

{\baselineskip=14pt \it 
The Abdus Salam ICTP, Strada Costiera 11, Trieste, Italy \\[1mm]}

\vspace{0.2in}
\end{center}

\vspace{0.2cm}
\begin{center}
{\bf Abstract} 
\end{center}
\vspace{0.2cm} 
We discuss the constraints on the inhomogeneous reheating scenario.
In particular, we discuss the prospects for low scale inflation with a
Hubble constant of the order of the gravitino mass, and the
possibility that an MSSM flat direction is responsible for the density
fluctuations. Thermal effects are generically fatal for the scenario,
and can only be avoided for small enough mass and couplings of the
flat direction field, excluding MSSM flat directions. Prompt decay at
the end of inflation bypasses thermal constraints, and is compatible
with both low scale inflation and MSSM flat directions.  However, the
level of non-Gaussianity is acceptably small only for a small window
of inflaton mass and couplings.  The upshot is that tuning of
parameters is needed for the inhomogeneous reheating scenario to work.

\newpage


\section{Introduction}

The inflationary paradigm in which the inflaton sector is responsible
for the density perturbations is economical.  One potential serves
many purposes: driving inflation, ending inflation, and generating the
observed spectrum of density perturbations.  The resulting models of
inflation are predictive, and falsifiable by experiment.  The flip
side of the coin, though, is that they are restrictive too, and it has
proven extremely hard to build realistic models of inflation. Often a
considerable amount of fine-tuning is needed for the model to satisfy
all constraints. For this reason it may be worthwhile to explore
alternatives.

Inflationary models can be ``liberated'' if its task list is
reduced~\cite{liberation}. This is the idea behind both the curvaton
scenario~\cite{early,curv} and the inhomogeneous reheating
scenario~\cite{FDR}, in which not the inflaton field but some other
field is responsible for the density perturbations.  The inflaton
sector merely serves to drive and end inflation, and is considerably
less constraint.  Of course, the price to be paid is that a new field
has to be introduced into the theory.  But the costs are minimized if
the new field and/or the new scales introduced are already present in
our models of particle physics.  There is already an extensive
literature exploring this possibility in the context of the curvature
scenario~\cite{curv_ext,curv_MSSM}, but little has been done in the
context of the inhomogeneous reheating scenario (but
see~\cite{FM,anupam}).  In this paper we will concentrate on the
inhomogeneous reheating scenario.  In particular, we will address the
following two questions. Is low scale inflation --- with the Hubble
scale during inflation of the order of the gravitino mass --- possible,
such that the inflaton sector can be naturally identified with SUSY
breaking sector?  Can any of the flat directions in the minimal
supersymmetric standard model (MSSM) be responsible for the density
fluctuations?

Low scale inflation is hard to realize in the conventional setting in
which the inflaton is responsible for the density fluctuations.~\footnote
{The exception are ``new inflation'' models, in which the inflaton
starts out close to the origin \cite{sarkar}.  When Taylor expanded
around the inflaton VEV, the linear term is anomolous small, making
this different behaviour possible.}
In that case the density fluctuations are proportional to $H_*/\mpl$,
with the subscript $*$ denoting the quantity evaluated at the time
observable scales leave the horizon, some 60 e-folds before the end of
inflation. Unless the slow roll parameter $\epsilon= (\mpl V'/V)^2 \ll
1$, which often requires some amount of fine tuning, high scale
inflation is needed to get the observed size of density perturbations.
In the curvaton scenario the density perturbations are proportional to
$H_* / \sigma_*$ with $\sigma$ the vacuum expectation value (VEV) of
the curvaton field.  Since the VEV can be much lower than the Planck
scale, low scale inflation is more natural in the curvaton
scenario. However, in~\cite{lyth} it was shown that for the scenario
to work the scale of inflation cannot be too low: $H_* \gtrsim 10^7
\GeV$.

In the inhomogeneous reheating scenario the density perturbations are
proportional to $H / S_*$ or $H / M$, depending on the form of the
decay rate, where $S$ is the VEV of the flat direction field
responsible for the density perturbations and $M$ is some cutoff
scale.  Since both $M$ and $S_*$ can be much smaller than the Planck
scale, also here, low scale inflation seems natural.  We will see,
that indeed, low scale inflation with $H_*$ of the order of the
gravitino mass is possible in this context, although only in a small
part of parameter space.  The fluctuating decay rate scenario can then
be married with any of the interesting inflaton models, in which the
inflaton originates from the SUSY breaking
sector~\cite{low_inflation}.

In both the curvaton and inhomogeneous reheating scenario a scalar
field other than the inflaton is responsible for the density
perturbations. This scalar has to be light compared with the scale of
inflation to be able to fluctuate freely, and produce the scale
invariant perturbation spectrum observed.  Obvious candidates for this
scalar field are the MSSM flat directions.  Within the curvaton
scenario this possibility has been studied in~\cite{curv_MSSM}.  It
was found that only under very special conditions can the MSSM scalar
produce the density perturbations.  The main problem is that the
curvaton has to (nearly) dominate the energy density before decay.
This means large initial VEVs and thus potential problems with
non-renormalizable operators, and this means late decay and thus
potential problems with disastrous finite temperature effects. In this
respect, the inhomogeneous reheating scenario offers better prospects,
as the MSSM flat direction does not need to dominate the energy
density at the time of decay.  Consequently, its VEV can be small
enough to avoid problems with the non-renormalizable potential, and
the density perturbations may be generated before thermal effects
become important.

This paper is organized as follows.  In the next section, we provide
the back ground material.  We describe those features of the inflaton
and MSSM sector important for the inhomogeneous reheating scenario.
In section \ref{S_constraints} we turn to the description of the
various constraints on the inhomogeneous reheating scenario.  Section
\ref{S_scales} discusses the various time scales in the problem,
giving further insights in the nature of especially the thermal
constraints. We address the prospects for model building,
concentrating on the possibility of low scale inflation and using MSSM
flat directions, for a polynomial decay rate in section \ref{S_pol},
while section \ref{S_const} discusses the same issues in the context
of an (approximate) constant decay rate.  In section \ref{S_psi} we
discuss the varying mass scenario, which is a variation of the
inhomogeneous reheating scenario, and point out its virtues and
drawbacks for model building.  Finally, we conclude in section
\ref{S_concl}.


\section{Preliminaries}

The idea behind the inhomogeneous reheating scenario is the following.
Consider the decay of the inflaton field, or more generally the field
dominating the energy density (FDED), into standard model degrees of
freedom.  Suppose now that the decay rate for this process depends on
the vacuum expectation value of some field. In supersymmetric theories
as well as in superstring inspired theories, the effective couplings
are functions of the various fields in the theory, and thus so is the
decay rate.  If one of these fields, call it $S$, is light during
inflation it can condense with a large VEV. Moreover, it fluctuates
freely during inflation.  As a result, the decay rate is spatially
fluctuating on superhorizon scales.  

FDED decay will happen at slightly different times in different parts
of the universe.  The regions in which decay has taken place are
filled with radiation, while the not yet decayed regions are matter
dominated.  The universe expands at a different rate in different
regions, resulting in fluctuations in the reheat temperature, hence, to
adiabatic density perturbations.

\subsection{Decay rates}

The decay rate of the field dominating the energy density can be
schematically written as $\Gamma = \lambda^2 m K$, with $m$ the FDED
mass, $\lambda$ its coupling, and $K$ a phase space factor.  All three
quantities can have a field dependence, leading to different
realizations of the inhomogeneous reheating scenario.  The various
possibilities will be discussed in detail in later sections.  For now
we just want to remark that we can divide the decay rates into two
general classes, which differ in their $S$-dependence.  Here and in
the following $S$ denotes the flat direction (``flaton'') field
responsible for the density perturbations.

The first class consists of decay rates polynomial in $S$, of the form
\be
\Gamma_\phi = \Gamma_0 S^p, \qquad (p \geq 1),
\label{P_gamma}
\ee
Decay rates of this form can arise from non-normalizable operators.
If there are several fields $S_i$ that can give rise to a decay rate
of the above form, the dominant contribution will come from the field
with the largest VEV.  A large VEV favors directions with a small mass.
This may naturally select a flaton with $m_S \lesssim 0.1 H_*$, leading
to a scale invariant spectrum.
 
The second class of decay rates are of the form
\be
\Gamma_\phi = \Gamma_0
\left[ 1 + \( \frac{S}{M}\)^q \right]^r
, \qquad (q \geq 1).
\label{C_gamma}
\ee
with $S < M$.  This is the $p=0$ limit of the polynomial decay
rate. At zeroth order the decay rate is constant, the $S$ dependence
comes only in through higher order terms. Such decay rates may be
obtained from normalizable operators, or from phase space effects.  We
will refer to the above decay rates as respectively ``polynomial'' and
(approximately) ``constant''.

\subsection{Inflaton sector}

We will not be concerned with the specific origin of inflation.  All
that is needed is that there is period of inflation, long enough to
solve the horizon and flatness problem, and with a Hubble constant
that is almost constant $|\dot{H_*}/H_*^2| \ll 1$.  However, the mass
$m_\phi$ of the field that stores most of the energy at the end of
inflation will be an important parameter, so we digress somewhat on
that.

In one-field models of inflation the inflaton mass is bounded by the
Hubble scale during inflation: $m_\phi \lesssim H_*$.  Since the
density perturbations are not produced by the inflaton field, the slow
roll parameter $\eta$ does not need to be much smaller than unity.
Fast roll inflation is possible for $m_\phi \gtrsim H_*$, but the
number of e-folds is negligible small unless $m_\phi \to H_*$.

There is more freedom in multiple field models of inflation.  Consider
for example hybrid inflation with a potential~\cite{hybrid}
\be
V(\phi,\chi) = V_0 + \delta V(\chi) - \frac{1}{2} m_\phi^2 \phi^2 
+ \frac{1}{2} \lambda' \chi^2 \phi^2 + \frac{1}{4} \lambda \phi^4.
\ee
In supersymmetric theories the couplings $\lambda'$ and $\lambda$ are
related. If the inflaton $\chi$ is responsible for the density
perturbations, $\lambda'$ has to be sufficiently small to assure a
scale invariant spectrum, but this requirement can be relaxed in the
inhomogeneous reheating scenario.  Inflation occurs in the regime $\chi
> \chi_c = m_\phi / \sqrt{\lambda'}$, and the inflaton rolls slowly in
the potential $\delta V(\chi)$.  Inflation ends when $\chi < \chi_c$,
and $\phi$ acquires a VEV
\be
\langle \phi\rangle = \frac{2 V_0^{1/2} }{m_\phi},
\qquad {\rm and} \qquad
\lambda = \frac{4 V_0}{\langle \phi \rangle^4} = 
\frac{m_\phi^4}{4 V_0},
\ee
where we have set $V = 0$ in the vacuum.  The field dominating the
energy density at the end of inflation is the waterfall field $\phi$.
Taking $\lambda \sim 1$, the mass of this field in the true vacuum is
$m_\phi \sim V_0^{1/4}$ and thus $m_\phi^2 \gg H_*^2 \sim
V_0$.~\footnote
{We will use units in which the reduced Planck mass $\mpl = 8\pi G
=1$}
Another example of two-field inflation is the recently proposed ``new
old inflation''~\cite{NOI}.  The potential is of the hybrid inflation
type, and also in this case $m_\phi \gg H_*$ is possible.

In the next sections the mass $m_\phi$ denotes the mass of the field
dominating the energy density at the end of inflation, i.e., the
inflaton field in one field models of inflation and the waterfall
field in hybrid inflation. With an abuse of language will refer to
this field in both cases as the inflaton field. We parametrize
\bea
m_\phi &\sim \beta H_*,  &\quad ({\rm 1FI}), \nonumber \\
m_\phi &\sim \beta \sqrt{H_*},  &\quad ({\rm 2FI}),
\label{mphi}
\eea
with $\beta \lesssim 1$ for both one field inflation (1FI) and two
field inflation (2FI) such as hybrid inflation.  Since it is not
expected that $m_\phi \gg V_0$, the $\beta$-factor is maximum for 2FI.

A variation of the inhomogeneous reheating scenario was proposed
in~\cite{FM}. Suppose that the inflaton decays into heavy particles
$\psi$, whose mass is set by a flat direction VEV.  The particles
freeze-out, and when they become non-relativistic they soon come to
dominate the energy density in the universe. The density perturbations
are generated during the decay of the $\psi$ particles, and are
sourced in this case by a varying mass as opposed to a varying
coupling.~\footnote
{Of course, one could also consider a decay rate for $\psi$ with a
varying coupling constant. This is the same as considering inflaton
decay with a varying coupling, only less economical, since an extra
field is introduced.}
This scenario seems less economical since it needs the introduction of
yet another field. We consider it for completeness though.  The mass
$m_\psi$ is bounded by the temperature $m_\psi \lesssim T$ for thermal
production. If they are direct inflaton decay products, $m_\psi <
m_\phi$ for perturbative decay, and $m_\psi < 10^2 -10^4 m_\phi$ for
bosons and fermions respectively in non-perturbative
decay~\cite{wimpzillas}.

We will only consider perturbative inflaton decay.  It is well known
that the waterfall field in hybrid inflation generically decays
non-perturbatively, in a rapid process dubbed instant
preheating~\cite{instant}.  The negative (mass$)^2$ at the end of
inflation leads to a spinodal instability, and the $\phi$ condensate
breaks up almost instantly. All long wavelength modes below some
critical wavelength $k_*$ are excited.  The waterfall field still
dominates the energy density, but now the energy is not only stored in
the zero-mode but in all modes with $k \lesssim k_*$.  If decay
happens before annihilation reactions become important, the
inhomogeneous reheating scenario still works as before.

\subsection{Low scale inflation} 
Several inflationary models have been constructed in which the
inflaton sector is linked to SUSY breaking~\cite{low_inflation}.  Then
naturally $V_0^{1/4} \sim 10^{-8}$ and $H_* \sim V_0^{1/2} \sim m_{3/2}
\sim 10^{-16}$ for gravity mediated SUSY breaking. In anomaly mediated
SUSY breaking schemes, the scale of inflation $V_0^{1/4}$ is larger by
one or two orders of magnitude, whereas for gauge mediation the scale
is smaller.

\subsection{MSSM flat directions}
The flat directions of the MSSM consist of gauge invariant operators
composed of MSSM scalar fields~\cite{gherghetta}. This polynomial is
commonly parametrized as $X = \phi^n$, with $n$ the dimension of
$X$. A non-zero VEV for $\phi$ will break the standard model gauge
symmetry. All fields entering in the flat directions which are left
after the Higgs mechanism (except the linear combination which
receives the VEV after diagonalization) have masses of order $h \phi$,
with $h$ a gauge coupling, due to their $D$-term couplings to the flat
direction VEV~\cite{ellis}.  In addition superpotential couplings of
the form $W = h \phi q q$ lead to effective mass terms $M_q \sim h
\phi$ with $h$ the MSSM Yukawas.

Therefore, if the flaton $S$ responsible for the density perturbations
is identified with one of the MSSM flatons $\phi$, at least one of its
couplings is of gauge strength $h \sim 0.1$.  The Yukawas vary between
$10^{-6} < h < 1$.


\section{Constraints}
\label{S_constraints}

We list here the various constraints on the inhomogeneous reheating
scenario.

\subsection{Density perturbations}

The density perturbations can be parametrized by the gauge invariant
quantity $\zeta$, which describes the the density perturbations on
uniform curvature slices~\cite{bardeen}.  The perturbation generated
by a fluctuating inflaton decay rate are~\cite{FDR,FM,perturbations}
\be
\zeta = \alpha \frac{\delta \Gamma_\phi}{\Gamma_\phi},
\label{dGamma}
\ee
with $\alpha$ the efficiency parameter. $\Gamma_\phi$ is to be
evaluated at the time of decay, when $H \sim \Gamma_\phi$.  After
reheating is completed the metric can be written as $d s^2 = - d t^2 +
g^2(\Gamma_\phi) t d x^2$~\cite{zaldarriaga}.  The function $g \propto
\Gamma_\phi^{- \alpha}$ parametrizes the difference in expansion rates
in different regions, and thus the resulting density perturbations.
It can be calculated numerically by integrating the coupled equations
of motion for the inflaton and the radiation bath, together with the
Friedman equation.  The slope gives the efficiency parameter $\alpha$,
which is shown in Fig.~1.

\begin{figure}[t]
\centering
\hspace*{-5.5mm}
\leavevmode\epsfysize=8cm \epsfbox{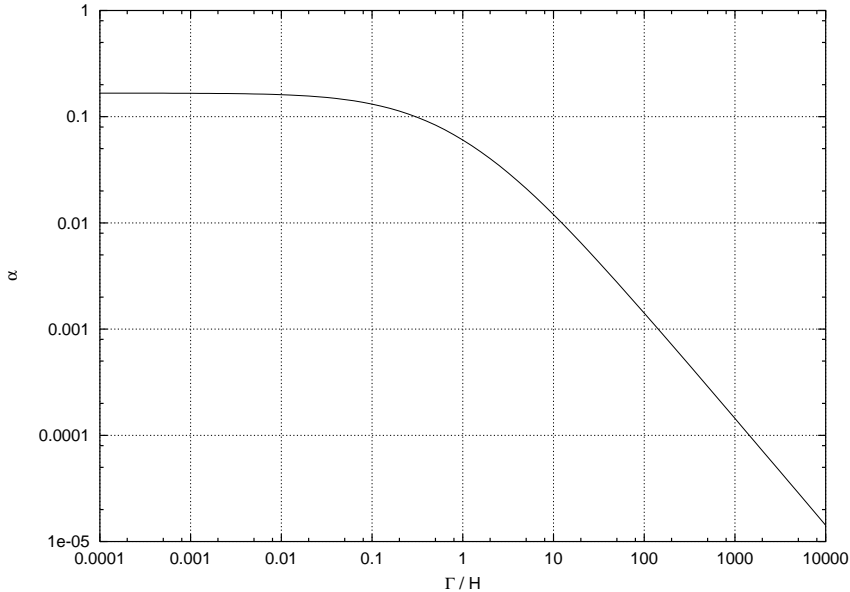}\\
\caption{Efficiency parameter $\alpha$ as a function of $\Gamma_\phi /
H_{\rm end}$}
\end{figure}

The parameter $\alpha$ depends on the ratio $\Gamma_\phi/H_{\rm end}$
with $H_{\rm end}$ the Hubble constant at the end of inflation.  If
$\Gamma_\phi \gg H_{\rm end}$, decay occurs almost instantaneously in
all regions of space, almost independently of the fluctuations in the
decay rate, and the resulting density perturbations will be small.  In
the opposite limit $\Gamma_\phi \ll H_{\rm end}$ the efficiency is
maximal, and $\alpha$ reaches its maximum value $\alpha = 1/6$, which
agrees with analytical estimates~\cite{FDR,FM,perturbations}.  We can
parametrize
\be
\alpha \approx \left\{
\begin{array}{lll}
&1/6,   & \qquad \Gamma_\phi \lesssim 10^{-2} H_{\rm end}, \\
& 1/6 - 1/15, & \qquad \Gamma_\phi \lesssim 10^{-2} - 1 H_{\rm end},\\
& 0.1 (H_{\rm end}/\Gamma_\phi), & \qquad \Gamma_\phi \gtrsim H_{\rm end}.
\end{array}
\right.
\label{alpha}
\ee

The value for $\zeta$ measured by WMAP is $ \zeta = 2 \times 10^{-5}
$~\cite{WMAP}.  The right amount of density perturbations for a
polynomial decay rate of the form Eq.~(\ref{P_gamma}) are obtained if
\be
\frac{\delta S}{S} \Big|_{\rm dec} = \frac{2 \times 10^{-5}}{\alpha p},
\label{P_S}
\ee
The equivalent for the constant decay rate of Eq.~(\ref{C_gamma}) is
\be
\frac{S^{q-1}\delta S}{M^q} \Big|_{\rm dec} 
= \frac{2 \times 10^{-5}}{\alpha q r},
\label{C_S}
\ee
where we have used $S/M \ll 1$.

If the density perturbations are produced during the decay by a field
$\psi$ other than the inflaton field, that dominates the energy
density sometime after the end of inflation, then generically
$\Gamma_\psi \ll H_{\rm end}$ and $\alpha = 1/6$.  If the FDED has a
varying mass, density perturbations are already produced before decay,
as domination happens at different times in different parts of the
universe.  The resulting perturbations after decay are~\footnote
{While finishing up this paper~\cite{vernizzi} appeared, claiming the
density perturbations produced in the fluctuating mass case is about a
factor $10$ larger than the estimate Eq.~(\ref{dGamma}), which is
taken from~\cite{FM}. The difference will not affect our conclusions
in an essential way.}
\be
\zeta = \frac{1}{6} \frac{ \delta \Gamma_\psi}{\Gamma_\psi} - 
\frac{1}{3} \frac{ \delta m_\psi}{m_\psi},
\ee
to be evaluated at the time of decay.  If the mass $m_\psi \propto S$,
the decay rate is polynomial and the density perturbations are up to a
minus sign given by Eq.~(\ref{P_S}) with $p=1$ and $\alpha =1/6$.

In the standard scenario in which the inflaton is responsible for the
density perturbations the non-Gaussianity is small.  The slow roll
conditions assure that the perturbations are Gaussian at production,
and as the adiabatic perturbations remain constant on superhorizon
scales, no further non-Gaussianity is produced.  In contrast, large
non-Gaussianity is possible in the inhomogeneous reheating
scenario. Apart from the possibility of non-Gaussianity at production,
non-Gaussianities can be produced due to the non-linear evolution if
$S$ starts rolling in the potential, due to the non-linear relation
between the decay rate and $\delta S$, and due to the non-linear
relation between the decay rate and the final
perturbations~\cite{FM,zaldarriaga,riotto}. The general rule is that
the less efficient the transfer of $S$ perturbations in metric
perturbations is, the larger the non-Gaussianities. This is easy to
understand, since if the transfer is less efficient, larger
fluctuations $\delta S / S$ are needed to obtain the observed
spectrum.

A level of Gaussianity as probed by WMAP is assured if 
\be
\delta S\lesssim 10 S,
\label{gaussian}
\ee
both at the time of production as well as at decay.  In particularly,
this means $\delta S_* \approx H_* /(2 \pi) \lesssim 10 S_*$.  The
transfer of perturbations is inefficient if the transfer parameter
$\alpha$ is small, or for the constant decay rate if the ratio $S/M$
is small.  Indeed, combining Eqs.~(\ref{P_S},~\ref{C_S}) with the
constraint from Gaussianity, Eq.~(\ref{gaussian}), bounds $\alpha
\gtrsim 10^{-4}$ and $\alpha (S_{\rm dec}/M)^{q} \gtrsim 10^{-4}$
respectively.  When $V''(S_*) \lesssim H^2$ the flaton zero mode and
its evolution starts rolling in its potential.  This may introduce
additional inefficiencies, which will be discussed in the next
subsection.

The produced spectrum of perturbations is nearly scale invariant, as
dictated by observations, if the mass of the fluctuating field is
sufficiently small $m_S \lesssim 0.1 H_*$. Note that the same must
hold for the effective mass $m_{\rm eff} = V''$ generated by higher
order terms in the potential.  A quartic term $V = \kappa |S|^4$ will
lift the flatness of the potential, unless $\kappa$ is exceedingly
small.  For this reason we will not consider such potentials in the
following. One should also consider non-renormalizable operators in
the potential of the form~\cite{DRT}
\be
V_{\rm NR} = \frac{|\kappa|^2 |S|^{2(n-1)}}{\Lambda^{2(n-3)}},
\label{VNR}
\ee
with $\Lambda$ some cutoff scale, e.g. the Planck mass. In an expanding
universe supersymmetry is broken dynamically, leading to soft mass
terms for all scalars of the order of the Hubble constant. Such large
masses $m \sim H$ will spoil the scale invariance of the
perturbations, and should be avoided. We will assume that the Hubble
induced flaton mass is at least one order of magnitude below its
canonical value, assuring it is unimportant at all times.  Soft mass
terms can be suppressed by either invoking symmetries, or by allowing
tuning.

The density perturbations produced trough the inhomogeneous reheating
mechanism are by assumption the dominant ones; other sources should
lead to negligible perturbations. In particular, the perturbations
produced by the inflaton field should be negligible small, which is
assured for $H_* \lesssim H_{\rm max} \equiv 10^{-6}$.  Further, at
the time of $S$-decay $\rho_S / \rho_{\rm total} \ll 1$, otherwise the
flaton perturbations will dominate through the curvaton mechanism. In
the curvaton scenario $\zeta \approx r/(2\pi) H_*/ S_*$, with $r =
\rho_S /\rho_{\rm total} \sim S_*^2 g^{-1}$ and $g$ the largest
flaton coupling, determining its decay rate. The curvaton
perturbations are sub-dominant if $g^ \gg S_*^2$.

\subsection{Evolution of $S$ and $\delta S$}

The decay rate is a function of the flat direction field, and
therefore a function of time.  The inflaton decays when $H /
\Gamma_\phi(S)$ drops below unity.  This excludes cases in which the
decay rate decreases faster than the Hubble constant.

Once the effective flaton mass becomes of the order of the Hubble
rate, $V''(S_*) \sim H^2$, the flaton starts oscillating in its
potential.  For a quadratic potential $\rho_S \sim a^{-3}$.  During
inflaton oscillations, the scale factor of the universe scales as $a
\propto H^{-2/3}$, and $S \propto H$. The ratio $H / \Gamma_\phi(S)$
remains constant, and the inflaton does not decay. If the potential is
dominated by non-renormalizable operators, then $S$ rolls down
approximately critically damped $V_{\rm NR}'' \sim H^2$.  We
parametrize the post-inflationary evolution of the flaton field
\be
S \equiv f S_*,
\label{f_def}
\ee
with
\bea
f \propto \left\{
\begin{array}{lll}
& H, & \qquad (V_{m_S}),\\
& H^{\textstyle \frac{1}{n-2}},  & \qquad (V_{\rm NR}),
\end{array}
\right.
\label{f}
\eea
where $V_{m_S} = (1/2) m_S^2 |S|^2$ and $V_{\rm NR}$ given by
Eq.~(\ref{VNR}). 

Further, the inflaton should decay before the decay of the flat
direction field: $\Gamma_\phi > \Gamma_S$.  This is automatically
satisfied if $\Gamma_\phi > m_S$ and flaton couplings less than unity.
The same constraints hold if it is not the inflaton but some other
field dominating the energy density, which has a varying decay rate.

As the Hubble constant drops below $\sqrt{V''(S)}$ not only the zero
mode but also the fluctuations $\delta S$ start evolving. The
equations of motion for the zero mode and the superhorizon
fluctuations are
\bea
&&\ddot{S} +3 H \dot{S} + V'(S) = 0, \nonumber \\
&&\ddot{\delta S} +3 H \dot{\delta S} + V''(S) \delta S= 0.
\label{eq_of_m}
\eea
For a quadratic potential, the equations of motion for $S$ and $\delta
S$ are identical (to linear order) and the ratio $(\delta S / S)$
remains constant.  If the potential is dominated by non-renormalizable
operators the fluctuations will be damped with respect to the
zero-mode according to~\cite{damping}
\be
D \propto  {H}^{\textstyle \frac{n-4}{2(n-2)}} ,
\label{D}
\ee
with the damping factor $D$ defined as
\be
\( \frac{\delta S}{S} \) \equiv D \( \frac{\delta S}{S} \)_*.
\label{D_def}
\ee
There is no damping for $n = 4$ operators.

Using the definitions of $f$ and $D$ we can express the equations for
the density perturbations Eqs.~(\ref{P_S},~\ref{C_S}) in terms of the
quantities during inflation:
\be
S_* \approx 8 \times 10^3 \alpha p D H_*,
\label{P_Sstar}
\ee
for a polynomial decay rate.  Here we have used $\delta S_* \approx
H_*/(2\pi)$.  Gaussianity as in Eq.~(\ref{gaussian}) requires $D
\gtrsim 10^{-3} p^{-1} ((1/6) /\alpha)$.  The equivalent expression
for a constant decay rate is
\be
\frac{H_* S_*^{q-1}}{M^q} \approx \frac{10^{-4}}{\alpha q r D f^q}
\label{C_Sstar}
\ee
with $Df^q \gtrsim 10^{-3}(M/S_*)^q ((1/6) /\alpha) (qr)^{-1}$ to
assure Gaussianity.  

Apart from the zero temperature potential the evolution of $S$ and
$\delta S$ can be sourced by finite temperature effects.

\subsection{Thermal constraints}

The radiation bath affects the flaton condensate in two ways, through
thermal scattering and through thermal corrections to the flaton
mass. The fields that couple to the flat directions through a
gauge/Yukawa coupling $h$, denote them by $\chi$, have a large
effective mass $m_\chi \sim h S$.  The $\chi$ particles are in thermal
equilibrium with the radiation bath, if their effective mass is
smaller than the temperature
\be
m_\chi \sim h S \lesssim T.
\label{cond_equil}
\ee
It is important to realize that even before inflaton decay has
completed, there is a dilute plasma with temperature~\cite{Tmax}
\be
T = (T_R^2 H)^{1/4} = (\Gamma_\phi H)^{1/4}, 
\label{T}
\ee
with $T_R \approx \sqrt{\Gamma_\phi}$ the reheat temperature at the
end of the reheating process when $\Gamma_\phi \sim H$.  In between the
end of inflation and inflaton decay the effective temperature scales
as $T \propto H^{1/4}$, instead of the $T \propto H^{1/2}$ during
radiation domination.  The plasma reaches its maximum temperature
$T_{\rm max}$ immediately after the end of inflation:
\be 
T_{\rm max} \sim \( \frac{H_{\rm end}}{ \Gamma_\phi} \)^{1/4} T_R
\sim (H_{\rm end} \Gamma_\phi)^{1/4},
\label{Tmax}
\ee
with $H_{\rm end}$ the Hubble constant at the end of inflation.

The back reaction of the $\chi$ quanta on the flat direction field
induces thermal corrections to the flaton mass.  If the $\chi$ quanta
are in thermal equilibrium, when Eq.~(\ref{cond_equil}) is satisfied,
there is an induced ``plasma mass'' of the form
\be
\delta m^2_{{\rm pl}} \sim h^2 \frac{n_\chi}{E_\chi} \sim h^2 T^2,
\qquad (hS < T),
\label{mth}
\ee
where the last equality hold if the $\chi$ quanta have a thermal
distribution with $n_\chi \sim T^3$ and $E_\chi \sim T$. There is also
a thermal correction in the opposite limit --- $T \lesssim m_\chi$,
and the $\chi$ quanta are out of equilibrium --- as a result of
integrating out the heavy degrees of freedom~\cite{dine,yanagida}. The
massless vector and chiral superfields, i.e., those not ``Higgsed'' by
the flat direction VEV $S$, generate a two-loop free energy
proportional to $h'^2 T^4$, with $h'$ the corresponding gauge or
Yukawa coupling.  The running of $h'$ changes at the renormalization
scale $\mu \sim m_\chi \sim h S$, if $m_\chi$ is charged under the
gauge group, respectively couples to the matter field in question.
Integrating out the heavy $\chi$ fields above this scale, leads to an
effective ``RG potential'' of the form:
\be
\delta V_{\rm RG} \sim c_T h'^4 T^4 \log \( \frac{hS}{T} \)^2,
\qquad (hS > T),
\label{Vth}
\ee
with $c_T$ a constant of order one which can have either positive or
negative sign, depending on the matter content. The sign is negative
if the running is dominated by integrating out Higgsed gauge bosons.

Large thermal masses, $m_{\rm th} > H > m_S$, induce early
oscillations of the flat direction field.  During these oscillation,
the energy density stored in the flat direction scales as~\cite{KT}
\be
\rho_S(H) = \frac{m(H)}{m(H_0)} \frac{H^2}{H_0^2} \rho_S(H_0).
\label{Sth}
\ee

Thermal evaporation of the flat direction through collisions with the
$\chi$ particles occurs if (1) the scattering rate is large
$\Gamma_{\rm scat} \gtrsim H$, (2) the $\chi$ particles are in thermal
equilibrium, and (3) the energy density stored in the thermal bath is
larger than the energy density in the flat direction. This last
condition will generically be satisfied.  The equilibrium condition is
given by Eq.~(\ref{cond_equil}).  The scattering rate is $\Gamma_{\rm
scat} \sim n_\chi \sigma$.  For a superpotential term $W = h S \chi
\chi$ the cross section for $\chi S$ scattering is $\sigma \sim h^2
\alpha / E_{\rm cm}^2$, where we have assumed scattering is dominated
by fermions. The typical center of mass energy is $E_{\rm cm} \sim
\sqrt{T m}$, the mean of the typical $\chi$ energy ($\sim T$) and $S$
energy ($\sim m = hT$, where the last equality holds when thermal
masses are important).  Then $\Gamma_{\rm scat} \sim h g^2 T$, with $g
\sim 0.1$ a gauge coupling~\cite{dine}.

Early thermal evaporation obviously kills the inhomogeneous reheating
scenario.  As we will see, early induced oscillations are generically
also fatal.  Therefore, for the fluctuating decay scenario to work,
thermal effects should be negligible.  There are three possibilities.
The first is that the particles coupling to the flat directions are
out of equilibrium.  Thermal scattering is negligible. There is a
renormalization group (RG) induced potential of the form
Eq~(\ref{Vth}). If the thermal mass is less than the Hubble scale
$\delta V_{\rm RG}'' \ll H^2$, the fields remains frozen and thermal
effects are negligible.

The second possibility is that the effective plasmon mass is lower
than the temperature. Thermal scattering is delayed with respect to
the onset of induced oscillations. Therefore, thermal effects are
absent it the induced plasma mass, Eq~(\ref{mth}), is smaller than the
Hubble rate.

The time when thermal effects become important may be delayed if the
thermal plasma is initially far from a thermal equilibrium
distribution.  If $n_\chi/E_\chi \ll T^2$, the plasma mass is then
below its equilibrium value.  The scattering rate may also be
suppressed, as it scales with $n_\chi$.~\footnote
{Note that there is an enhancement in the scattering rate for smaller
induced mass, as $\sigma \propto 1 / m_{\rm th}$.}
Likewise, if the $\chi$ are non-relativistic, the RG potential can be
suppressed if the free energy of the massless field are below their
equilibrium value.  

An initial distribution with $n_\chi/E_\chi \ll T^2$ can happen if the
plasmons are direct inflaton decay products with $E_\chi \sim m_\phi
\gg T$.  Or else, if the plasmons are not inflaton decay products,
their initial number density is small.  The plasmons acquire chemical
equilibrium when the rate for number changing interactions
$\Gamma_{\rm int}$ becomes of the order of the Hubble constant.
$\Gamma_{\rm int} \sim h_{\rm pl}^2 g^2 T$, where $h_{\rm pl}$ is the
coupling of $\chi$ to the thermal bath and $g$ a gauge coupling. If
the plasmons carry gauge charges, $h_{\rm pl} \sim g \sim 10^{-1}$;
the time scale for chemical equilibrium $\Gamma_{\rm int} \sim
10^{-4}T$ is smaller than the plasma mass $hT$, and the onset of
plasma induced oscillations can be delayed only for $h \gtrsim
10^{-4}$.\\

The induced thermal masses and scattering rates are a function of the
temperature $T \sim (H \Gamma_\phi(S))^{1/4}$. Hence, the constraints
depend on how the inflaton decay rate depends on $S$, and are
therefore model dependent. The dependence of the plasma temperature on
$H$, Eq.~(\ref{T}), is derived for a constant inflaton decay rate. We
will approximate the temperature by this formula also in the case of a
variable decay rate, so that the temperature becomes a function of
both $H$ and $S$.

If oscillations of the flat direction field are set off by thermally
induced masses, the field amplitude decreases as given by
Eq.~(\ref{Sth}).  For plasma masses $m_{\rm pl} \sim hT$ this implies
\be
f \propto H^{7/(8+p)}, 
\label{f_mth}
\ee
with $p=0$ for a constant decay rate.  Since the potential is
approximately quadratic $D \sim 1$.  We postpone a discussion of $f$
and $D$ for RG induced potentials to the next section.\\

To avoid gravitino overproduction and thereby spoiling the successful
nucleosynthesis predictions, the reheat temperature has to be
sufficiently small $T_R \lesssim 10^{-10}$, or equivalently $\Gamma
\lesssim 10^{-19}$~\cite{gravitino}.  This is not a hard bound, in the
sense that depending on the specifics of the SUSY breaking mechanism,
the bound can shift some orders of magnitude.  Moreover, the gravitino
problem can be solved for example by invoking a period of thermal
inflation~\cite{thermal_inf}. A bound that cannot be tampered with is
the big bang nucleosynthesis (BBN) bound: The field dominating the
energy density should decay before BBN, $T_R \gtrsim 10^{-22}$ or
$\Gamma \gtrsim 10^{-43}$.

In this context, we should also remark that the baryon asymmetry of
the universe has to be created either during or after FDED decay. If
FDED decay happens below the electroweak scale $\Gamma_\phi \lesssim
10^{-33}$, sphalerons are out of equilibrium, and an Affleck-Dine like
mechanism for baryogenesis is required.


\section{Time scales} 
\label{S_scales}

The important time scales in the problem are the following.  Inflation
ends at
\be
H_{\rm end} \sim \zeta H_*,
\label{Hend}
\ee
with $\zeta$ parameterizing the difference between the time observable
scales leave the horizon, and the end of inflation.  For simplicity we
set $\zeta \sim 1$, although lower values are possible.  For example,
in 1FI with a quartic potential $\zeta \sim 0.1$.  

The large VEV of the inflaton gives a large mass to its decay
products, and inflaton decay is kinematically forbidden during
inflation.  Decay becomes possible as soon as the inflaton starts
oscillating in its potential.  Therefore, the Hubble constant at the
time of decay is $H_{\rm dec} \sim \min[\Gamma_\phi, \,H_{\rm end}]$.

The scales at which the vacuum mass and effective mass generated by
non-renormalizable operators become important are respectively $H \sim
m_S$ and
\be
H_{\rm NR} \sim \frac{\kappa S^{n-2}}{\Lambda^{n-3}}.
\label{HNR}
\ee

We denote the scale at which thermal effects become important by
$H_{\rm th}$. The particles coupling to the flat direction are in
equilibrium for $T > hS $, or equivalently $H >H_{\rm eq}$, with
\be
H_{\rm eq} \sim \min \left[
\frac{h^4 S^{4}}{\Gamma_\phi},\,H_{\rm end} \right].
\label{Heq}
\ee
Then for $H >H_{\rm eq}$, $H_{\rm th} \sim H_{\rm pl}$ the time scale
at which the plasma mass of Eq.~(\ref{mth}) becomes of the order of
the Hubble scale, while for $H < H_{\rm eq}$, $H_{\rm th} \sim H_{\rm
RG}$ the time scale at which the RG potential of Eq.~(\ref{Vth})
becomes important.  Here
\bea 
H_{\rm pl} & \sim& 
\min \left[h^{4/3} \Gamma_\phi^{1/3}, \,H_{\rm end} \right], 
\label{Hpl} \\ 
H_{\rm RG} &\sim&
\min \left[\frac{h^4 \Gamma_\phi}{ S^{2}} 
\( 2p+1 + p(p-1) {\log(\frac{hS}{T})}\) ,\,H_{\rm end}
\right], 
\label{HRG}
\eea
where $p=0$ corresponds to the constant decay rate. As discussed in
section \ref{S_constraints}, evaporation is delayed with respect
to the onset of thermally induced oscillations.  The onset of thermal
effects can be delayed if the initial plasma is far from an
equilibrium distribution, and the number densities of the particles
giving mass to $S$ is less than the equilibrium value $n_\chi \ll
T^3$.  The delay is until number changing interactions become
important, at $H_{\rm del} \sim \Gamma_{\rm int}$ with
\be
H_{\rm del} \sim
\min \left[(h_{\rm pl} g^2)^{4/3} \Gamma_\phi^{1/3}, \,H_{\rm end} \right], 
\label{Hdel}  
\ee
with $h_{\rm pl}$ the coupling of the particle responsible for the
induced thermal mass of $S$, i.e., the plasmon $\chi$ which couples to
$S$ if the $\chi$'s are relativistic, and the light degrees of freedom
whose free energy is altered by the running of $\chi$ for
non-relativistic $\chi$'s. \\

We can then distinguish the following cases:

\begin{enumerate}

\item

$\Gamma_\phi \gtrsim H_{\rm end}$.  Inflaton decay is kinematically
inaccessible during inflation, and the inflaton decays promptly at the
end of inflation.  

\item

The inflaton decays after the end of inflation, but while the flaton
is still frozen, i.e., before the onset of oscillations of the flat
direction field.  Consequently, there is no damping $D=1$ and $S_{\rm
dec} = S_*$.  This is the case for
\be
H_{\rm end} > \Gamma_\phi > m_S, \, H_{\rm th}, H_{\rm NR}.
\ee

\item

Inflaton decay occurs after the end of inflation, and after the flaton
field has started rolling in the quadratic potential, the
non-renormalizable potential and/or the thermally induced
potential. Damping and the evolution of $S$ should be taken into
account, and $f$ and $D$ are generically different from unity.

\end{enumerate}

\paragraph{Case 1}

Case 1 --- prompt decay --- is obtained if $\Gamma_\phi \gtrsim H_{\rm
end}$. The transfer of flaton perturbations to metric perturbations is
inefficient in this limit, and large non-Gaussianity is possible.
Gaussianity as in Eq.~(\ref{gaussian}) and the requirement of prompt
decay together constrain the decay rate, or equivalently the parameter
$\Gamma_0$ (see Eqs.~(\ref{P_gamma},~\ref{C_gamma})), to the range
\bea
(5p \times 10^2 H_*)^{-p}  \lesssim & 
{\displaystyle \frac{\Gamma_0}{H_*}}
& \lesssim {5p \times 10^2} H_*^{-p},
\hspace{1.cm} \qquad (\Gamma_\phi \; {\rm pol.}) 
\label{P_prompt}
\\
1 \lesssim & {\displaystyle \frac{\Gamma_0}{H_*} }
& \lesssim {5qr \times 10^2} \( \frac{H_*}{M} \)^{q},
\qquad \,(\Gamma_\phi \; {\rm cnst.}) 
\label{C_prompt}
\eea
for a polynomial and constant decay rate respectively.  Here we have
used Eqs.~(\ref{P_Sstar},~\ref{C_Sstar}). There is only a small window
of $\Gamma_0$ values for which prompt decay is possible.

There are no thermal or damping effects, and the only constraints come
from the observed magnitude of perturbations
Eqs.~(\ref{P_Sstar},~\ref{C_Sstar}) with $f = D =1$, and scale
invariance $m_S, \, \sqrt{V_{\rm NR}''(S_*)} \lesssim 0.1 H_*$.  In
addition the gravitino problem has to be addressed for $H_* \gtrsim
10^{-19}$.

\paragraph{Case 2}

Inflaton decay occurs after the end of inflation, but before the
Hubble constant drops below the effective mass of the flat direction
field, which aside from the vacuum mass can have contributions from
thermal effects and from non-renormalizable operators in the
Lagrangian. The zero-mode and fluctuations are frozen, and $f = D =1$.

The flaton vacuum mass $m_S$, respectively the effective mass from
non-renormalizable operators, is smaller than the Hubble constant at
inflaton decay for
\be
H_{\rm dec} \gtrsim m_S,\, H_{\rm NR}.
\label{NR_mass}
\ee

Thermal effects play no r\^ole if the plasmons are in equilibrium, but
inflaton decay happens before the plasma mass becomes important, that
is if
\be
H_{\rm dec} > H_{\rm pl}, \,H_{\rm eq} 
\label{pl_bnd}
\ee
It should be understood that $ H_{\rm pl}, H_{\rm eq} < H_{\rm end}$,
i.e., plasma masses are not important immediately after inflation has
ended, nor are the plasmons out of equilibrium from the start.
Another possibility for the thermal effects to be absent is that the
plasmons fall out of equilibrium before the equilibrium thermal
effects become important.  This includes the case in which the
plasmons are out of equilibrium right from the start.  A RG correction
to the potential is generated, which should be negligible $H_{\rm RG}
< H_{\rm dec}$.  Both conditions are satisfied for
\bea
H_{\rm eq} &>& H_{\rm pl}, \, H_{\rm dec} ,
\nonumber \\
h &\lesssim& S^{1/2}.
\label{RG_bnd}
\eea
Finally, thermal effects may not set in until $H \sim H_{\rm del}$, if
the plasma is initially far from the thermal equilibrium distribution.
Thermal effects are avoided if
\be
H_{\rm del} \lesssim H_{\rm dec}
\label{delay}
\ee

\paragraph{Case 3} 

The flaton starts rolling in its potential before inflaton decay.  
A polynomial decay rate changes with changing $S$, whereas a constant
decay rate remains (approximately) constant.  Hence, these two cases
are quite different and we will discuss them separately.\\

As said, the decay rate is time-dependent for a polynomial decay
rate. The first remark to be made is that the expression for the
density perturbations Eq.~(\ref{dGamma}) is derived for a constant
decay rate.  However, we expect that the time-dependence will not
significantly alter the size of the final perturbations.

Since the decay rate changes with time, there is the possibility that
the decay rate decreases faster than the Hubble constant, and inflaton
decay will not occur.  In fact, this is the case if the flaton starts
oscillating in the quadratic potential, when $H \lesssim m_S$.  The
flaton red shifts $S \propto H$, see Eq.~(\ref{f}), and decay is
impossible for all $p \geq 1$. Similarly, decay does not occur when
the flaton starts rolling in the non-renormalizable potential, and $n
\leq p+2$.  If oscillations of the flat direction field are set off by
a plasma mass $m_{\rm pl} \sim h T$, the field amplitude decreases as
given by Eq.~(\ref{f_mth}).  For $p \geq 2$ the decay rate decreases
faster than the Hubble rate, and the inflaton cannot decay during
thermally induced oscillations either.  Therefore
\be
\Gamma_\phi > m_S,\, H_{\rm th,\,p \geq 2}^{\rm eq},
\, H_{\rm NR,\,p \geq n+2} 
\label{P_nodecay}
\ee
should be satisfied for the inhomogeneous reheating scenario to be
possible.  Here the superscript eq indicates that the plasmons
coupling to the flat direction are in thermal equilibrium, and the
subscripts $p \geq 2$ and $p \geq n+2$ indicate that only decay rates
satisfying these conditions are implied.

For the same Hubble constant during inflation, the flat direction VEV
at inflaton decay is smaller (larger) than in case 1-2, due to both
the damping of the fluctuations and the evolution of $S$.  Now $S =
fS_* = 8\times 10^3 \alpha p fDH_*$, with $\alpha \approx 1/6$. For
$fD < 1\; (FD >1)$, inflaton decay occurs for Hubble constants which
are a factor $(fD)^p$ smaller (larger) compared to case 2. The bounds
on the flaton mass and couplings,
Eqs.~(\ref{NR_mass},~\ref{pl_bnd},~\ref{RG_bnd},~\ref{P_nodecay}), are
stronger (weaker) by factors of $(fD)^p$.  In addition, there is the
new constraint that the damping should not be too large, $D \gtrsim
10^{-3}$, to assure a Gaussian perturbation spectrum.

Consider first the effect of the non-renormalizable potential $V_{\rm
NR}$ of Eq.~(\ref{VNR}) with $n > p+2$, so that inflaton decay is
possible. If $H_{\rm NR} > \Gamma_\phi$, the flat direction field
starts slow rolling in the potential before the epoch of inflaton
decay.  For simplicity, we will concentrate on the case that thermal
effects are negligible at all times $H_{\rm NR} > \Gamma_\phi > m_S,
H_{\rm th}$. Then $fD <1$ and all constraints are stronger. The decay
rate at $H < H_{\rm NR}$ is
\be
\Gamma_\phi(H) =  \( \frac{H}  {H_{\rm NR}} \)^{2/(n-2)} 
\Gamma_\phi(H_{\rm NR})
\ee
Decay occurs at $\Gamma_\phi(H_{\rm dec}) \sim H_{\rm dec}$.  The
damping factor is $ D = ( H_{\rm dec}/{H_{\rm NR}}
)^{(n-4)/2(n-2)}$. The amplitude $S_*$ can be found as a function of
$H_*$ by solving the equation $S_* \approx 10^3 pD H_*$.  The
amplitude at the time of inflaton decay is $S_{\rm dec} \sim (H_{\rm
dec} / H_{\rm NR})^{1/(n-2)} S_*$.  For example, for $p=2$ and taking
$\kappa = 1$ and $\Lambda = 1$, this leads to
\bea
S_* &\sim&  (10^6 \Gamma_0 H_*^2)^{1/(n-2)},
\nonumber \\
\frac{H_{\rm dec}}{H_{\rm NR}} &\sim&  
\frac{10^{-6} \Gamma_0^{2/(n-4)}}{H_*^2}. 
\eea

The RG induced potential becomes important before inflaton decay if
\be
H_{\rm eq} > H_{\rm pl}, \, H_{\rm dec}, 
\qquad \& \qquad
h \gtrsim S^{1/2}.
\label{RG_yes}
\ee
The behavior of the zero mode and the fluctuations depends critically
on the sign of $c_T$.  The equations of motions are
\bea
\ddot{S} + 3 H \dot{S} 
+ {A H} {S^{p-2}} \(p \log(\frac{hS}{T}) + 1\) S = 0, \nonumber \\
\ddot{\delta S} + 3 H \dot{\delta S} 
+ {A H}{S^{p-2}} \(2p-1 + p(p-1) \log(\frac{hS}{T})\) \delta S = 0,
\label{P_eom}
\eea
with $A =c_T 2 h^4 \Gamma_0$, which can have either sign depending on
the sign of $c_T$.  For positive sign $c_T >0$ and $p \geq 2$ the
effective potential is up to factors of $\log(S)$ given by $V_{\rm th}
\propto S^p$. The field rolls towards lower values, and $f$ and $D$
are approximately as given in Eqs.~(\ref{f},~\ref{D}).  The thermal
potential shuts off when $h S \lesssim H$, at which point a plasma
mass $hT$ is generated.  If $H_{\rm pl} < H$ the field freezes, and as
the temperature drops the plasmons again fall out of equilibrium.  The
induced RG potential leads to the decrease of $S$ until the plasmons
regain equilibrium. And so forth.  As a result, the VEV tracks $h S
\sim T$.  Inflaton decay should happen before the flaton mass or
plasma effects become important, i.e.,
Eqs.~(\ref{NR_mass},~\ref{pl_bnd}) should be satisfied.  If the
tracking is halted due to the effects of the non-renormalizable
potential, $S$ decreases further, but the inhomogeneous reheating
scenario is still possible for $n>p+2$. Both the effects of $V_{\rm
NR}$ and $\delta V$ with $c_T > 0$ lead to $fD < 1$ and all bounds are
stronger compared to case 2.

The situation is completely different for $c_T <0$. Now the potential
is minimized for large $S$. When damping is negligible, for $p \geq 2$
the instability leads to exponential growth of the zero mode. For $p
\geq3$ the growth is stopped by damping, when $H_{\rm RG} \sim H^2$,
and the VEV tracks $S^{p-2} \sim H/A$. For $p=2$ however, the thermal
mass is up to log factors independent of $S$ and damping remains
unimportant. The growth of the zero mode is halted instead when non-renormalizable
terms become important, or when the inflaton decays, whatever comes
first.  If the log term in Eq.~(\ref{P_eom}) does not already dominate
initially, with the the fast growth of $S$ it will soon do so.  Then
for $p=2$ the zero mode and the fluctuations roll in the same
effective potential and their ratio remains constant, $D \sim
1$. Since $fD \gg 1$, inflaton decay is earlier making constraints
weaker.  Moreover, since $S$ increases, the out of equilibrium
condition $hS > T$ remains valid. \\

For an approximately constant decay rate of the form
Eq.~(\ref{C_gamma}) the time evolution of the flaton will not alter
the time of decay.  Inflaton decay can still occur, for example, after
the flaton starts oscillating in its quadratic potential.

For the same Hubble constant during inflation, the flat direction VEV
at inflaton decay is smaller than in case 1-2, due to both the damping
of the fluctuations and the evolution of $S$, see Eq.~(\ref{C_Sstar}).
The decay rate does not alter, but the bounds on $H_*,\,S_*$ and $M$
to get the right density perturbation are stronger for $f^q D < 1$.
Specifically, taking Gaussianity and the requirement $S < M$ into
account, constrains $f^q D \gtrsim 10^{-2}$.  For oscillations set off
by the mass term, the non-renormalizable potential, or by plasma
masses $f^q D < 1$.  From the expressions for $f$ and $D$,
Eqs.~(\ref{f},~\ref{D},~\ref{f_mth}), it follows that
\be 
m_S,\, H_{\rm pl} \lesssim 10^2 \Gamma_\phi
\quad \& \quad
H_{\rm NR} \lesssim 10^4 \Gamma_\phi
\label{notmuch}
\ee
for the varying decay rate scenario to work.  The flaton VEV and $H_*$
should be tuned even more than for cases 1 and 2.

Things may be different for the RG potential of Eq.~(\ref{Vth}). For a
constant decay rate, the zero mode and fluctuations have mass terms
with opposite sign
\bea
\ddot{S} + 3 H \dot{S} + \frac{A H}{S^2} S = 0, \nonumber \\
\ddot{\delta S} + 3 H \dot{\delta S} 
- \frac{A H}{S^2} \delta S = 0
\eea
with $A = c_T 2 h^4 \Gamma_0$. For $c_T > 0 $ the zero mode decreases,
while the fluctuations increase, and vise versa for $c_T <0$.  For
$c_T >0$, the decrease of $S$ is halted when the plasmons fall out of
equilibrium, and $hS$ tracks $T$. The interesting case is $q=1$, since
only then $f^q D > 1$. For $c_T >0$ the growth of the zero-mode is
halted by damping, and $\delta V'' \sim H^2$. Only for $q \geq 3$ is
$f^qD \propto H^{-(q-2)/2} > 1$.

However, there is not much that can be done to improve the situation.
This can easily seen by looking at the expression for the density
perturbations at the time of decay, Eq.~(\ref{C_S}), which implies
$\delta S_{\rm dec} > 10^{-4} M$.  Gaussianity and the cutoff
constrain $10 \delta S_{\rm dec} < S_{\rm dec} < M$.  For a given
cutoff and Hubble scale during inflation, there is only a small window
for $S_{\rm dec}$.  For $c_T > 0$, $S$ increases towards earlier time,
but it cannot excess the cutoff scale.  For $c_T < 0$, $S$ decreases
towards earlier time whereas $\delta S$ increases, leading to large
non-Gaussianities when $S \sim \delta S$.  There is no room for $S$ to
either decrease or increase much, and a considerable amount of tuning
is needed if the RG potential is to dominate at some point.


\section{Inflaton decay through non-renormalizable operators}
\label{S_pol}

A polynomial decay rate, as in Eq.~(\ref{P_gamma}), can arise from
non-renormalizable superpotential couplings of the form~\cite{FDR}
\be 
W = \lambda_0 \frac{q_i}{M} \phi q_j q_k 
\label{NR_op}
\ee 
with $M$ the cutoff scale, and $S = \langle q_i \rangle \neq 0$ is the
field responsible for the density perturbations.  This corresponds to
5-dim non-renormalizable operators in the potential. Such operators
may be obtained from integrating out physics above the cutoff scale
$M$. For inflaton decay into MSSM degrees of freedom the combination
$q_i q_j q_k$ should form a gauge invariant.  For $S$ part of a MSSM
flat direction, the only combinations are $q q q$, $\bar{q} \bar{q}
\bar{q}$ and $ h q \bar{q}$ with $q$ and $\bar{q}$ left-handed
quark/lepton superfields and their charge conjugate. Inflaton decay is
unsuppressed for $m_\phi \gtrsim m_q$, with $m_q$ the lightest
quark/lepton fields in the superfields $q_j, q_k$.  For $S$ a SM
singlet, the only possibility is $S h_u h_d$~\footnote
{In this case a normalizable superpotential coupling $W = \lambda' \phi
q_j q_k$ is also possible.  The decay through non-renormalizable
operators dominates for $\lambda_0 S / M > \lambda'$.}

The effective coupling is $\lambda = \lambda_0 (S /M)$.  Then the
decay rate is of the form Eq.~(\ref{P_gamma}) with~\footnote
{Higher values of $p$ can be obtained through interactions $W =
\lambda_0 (S/M)^m \phi q_j q_k$ with $m \geq 2$.  Since the
constraints are stronger for higher values of $p$ we will not discuss
this possibility.}
\be 
\Gamma_0 =  \frac{\lambda_0^2}{8\pi M^2} {m_\phi}, 
\qquad p =2.
\label{NR_gamma}
\ee 
with $m_\phi = \beta H_* \,(\beta \sqrt{H_*})$ for 1FI (2FI), and
$\beta \lesssim 1$.  The value $\Gamma_0$ is bounded from above by
$\lambda_0, \, \beta \lesssim 1$.  The lower bound comes from the
requirement that the inflaton decays before BBN:
\be 
\Gamma_0 \gtrsim \frac{10^{-47}}{ (H_* f D \alpha)^2}. 
\label{NR_BBN}
\ee

\subsection{Low scale inflation} 
Is the fluctuating decay scenario with a decay rate of the form
Eq.~(\ref{P_gamma},~\ref{NR_gamma}) compatible with low scale
inflation $H_* \sim m_{3/2} \sim 10^{-16}$?  And if so, what are the
bounds on the couplings and mass of the flaton field?

The decay rate depends on the specifics of the inflaton sector.
However, $\Gamma_0$ can be bounded by BBN and the scale of inflation
to lie in the range $10^{-17}/(fD\alpha)^2 \lesssim \Gamma_0 \sim
10^{-18} (10^{-10}) \beta \lambda_0^2/M^2$ for 1FI (2FI). For $M \sim
1$ there is no parameter space for 1FI. The cutoff should be larger
than the flat direction VEV $M > S_* \approx 10^{-12} \alpha D
H_*$. The gravitino problem is absent for $\Gamma_0 \lesssim 10^{7}/
(fD)^2$.

\paragraph{Case 1}
Prompt decay is possible for $10^{10} \lesssim \Gamma_0 \lesssim
10^{19}$, as follows from Eqs.~(\ref{P_prompt}). Such large decay
rates are not possible in 1FI, while for 2FI it can be obtained for
$10^{20} < \beta \lambda_0^2/M^2 < 10^{29}$.  This means $10^{-14}
\lesssim M \lesssim 10^{-10}$ and $\lambda_0^2 \beta \gtrsim
10^{-4}$. The scale $M$ cannot be identified with the scale of
inflation, and a new scale has to be introduced in the theory.

The flat direction couplings are unconstrained, whereas its mass $m_S
\lesssim 10^{-17}$, in order to obtain a scale invariant spectrum.  The
effective mass generated by the non-renormalizable potential is also
sufficiently small for Planck suppressed operators with $\kappa, \,
\Lambda \sim 1$. Note however, that if these operators are suppressed
by the same scale as the effective inflation coupling, i.e., $\Lambda
\sim M$, then $n=4$ operators should be absent or suppressed $\kappa
\ll 1$.  Further, $M \gtrsim 10^{-11} \, (10^{-12})$ for $n =5 \,
(n=6)$ operators.

The gravitino problem has to be addressed.

\paragraph{Case 2}
For smaller decay rates, $\Gamma_0 \lesssim 10^{10} D^2$, the inflaton
decays well after the end of inflation and case 2 and 3 apply. There
are additional constraints, of which the strongest ones comes from
finite temperature effects: early thermally induced oscillations and
thermal evaporation.  The inflaton decays while the flat direction
field is still frozen, and case 2 applies, if these thermal effects
are negligible small.

A word of caution here.  All estimates are order of magnitude
estimates.  But this is even more so when thermal effects are
considered.  The reason is that the thermally induced mass by
particles in (out of) equilibrium is a good approximation in the limit
$hS \ll T \,(hS \gg T)$.  However, in the limit $hS \to T$ both
approximations break down.

The first possibility for thermal effects to be absent is that the
particles coupling to $S$ fall out of equilibrium before plasma
effects become important, and with the RG potential small
Eq.~(\ref{RG_bnd}). This is only possible for small rates $\Gamma_0$,
inconsistent with the BBN bound.

The second possibility is that the plasmons are in equilibrium until
inflaton decay, and the plasma mass is small, Eq.~(\ref{pl_bnd}). This
requires $h \lesssim 10^{-13} \Gamma_0^{1/2}$.  Taking $M \sim 1$,
this implies very small couplings and mass: $h \lesssim 10^{-17}$ and
$m_S \lesssim 10^{-36}$ for 2FI. Identifying the cutoff with the scale
of inflation $M \sim \sqrt{H_*} \sim 10^{-8}$ improves the situation
considerably, but still $h \lesssim 10^{-19} \, (10^{-16})$ and $m_S
\lesssim 10^{-20} \, (10^{-28})$ for 2FI (1FI).  The constraints are
weakest in the limit $M \to 10^{-13}$, but at the expense of
introducing a new scale in the system, and of gravitino over
production.

Finally, there is the possibility that the initial plasma is far from
an equilibrium distribution, and the onset of thermal effects is
delayed, Eq.~({\ref{delay}).  $H_{\rm del} \lesssim \Gamma_\phi$ gives
the same constraints as in the previous paragraph, with the
replacement $h \to 10^{-2} h_{\rm pl}^2$.  Here $h_{\rm pl}$ is the
coupling of the $\chi$ particles (those particles coupling to $S$) to
the plasma.  If $\chi$ has gauge charges $h_{\rm pl} \sim 0.1$, and
delay of thermal effects is not possible.

The non-renormalizable operators are sub-dominant until inflaton
decay, and Eq.~(\ref{NR_mass}) is satisfied, for $ \Gamma_0 >
10^{-13(n-4)} (\kappa/\Lambda)^{n-3}$.  For $M, \, \Lambda \sim 1$ and
2FI $n=4$ operators should be suppressed.  For $M \sim 10^{-8}$ and
$\Lambda \sim 1$ non-renormalizable operators are consistent with 2FI,
and with 1FI for $n \geq 5$.  However, if $M \sim \Lambda \sim
10^{-8}$ one needs $n \geq 5 \, (n \geq 6)$ for 2FI (1FI).

\paragraph{Case 3}

Case 3 applies if the flat direction field starts rolling in its
potential before inflaton decay.  If the potential is dominated by the
mass term $m_S$, the plasma mass $m_{\rm pl} \sim hT$, or $n=4$
non-renormalizable operators the decay rate decreases faster than the
Hubble constant, and decay cannot occur. Non-renormalizable operators
with $n \geq 5$, and RG induced potentials with $c_T >0$ have $fD < 1$
due to evolution of the zero-mode and its fluctuations.  The already
tight constraints of case 2 become stronger by appropriate factors of
$fD$.

The only possibly interesting case is when the potential is dominated
by RG effects with $c_T < 0$, see Eq.~(\ref{Vth}). Both the zero-mode
and its fluctuations grow exponentially, so that $f \gg 1$ and $D \sim
1$, and hence $fD \gg 1$.  A RG potential is generated when (see
Eq.~(\ref{RG_yes}))
\be
h \gtrsim \min[ 10^{13/2} \sqrt{\Gamma_0},\, 10^{5/2} (\Gamma_0)^{1/4}],
\label{RG}
\ee
and therefore $\Gamma_0 \lesssim 10^{-10}$ for couplings less than
one. Moreover, if $h > S^{1/2} \sim 10^{-13/2}$ the induced thermal
mass is larger than the decay width, and the inflaton start rolling in
$\delta V_{\rm th}$ before decay.  The small rates $\Gamma_0$ required
are naturally for $M \sim 1$.  A large cutoff has the additional
advantage that the zero mode can grow by a huge amount without
exceeding the cutoff. Note also, that small $\Gamma_0$ is consistent
with the BBN bound for $f \gg 1$.

For all $\Gamma_0$ values under consideration~\footnote
{For $10^{-16} < \Gamma_0 < 10^{-10}$ the growth of the zero mode can
start immediately at the end of inflation, at $H \sim H_{\rm
end}$. The non-renormalizable potential becomes important before decay
for all $n$.}
, the zero-mode starts growing rapidly as soon as $H \sim H_{\rm RG}
\sim \Gamma_0 h^4$, until its growth is halted by either the
non-renormalizable potential ($\delta V_{\rm RG} \sim V_{\rm NR}$), or
inflaton decay ($\Gamma_\phi \sim H$).  Approximating the increase to
be instantaneous, the maximum value of the zero mode before the
non-renormalizable potential becomes important is $S_{\rm max}^{n-2}
\sim H_{\rm RG} \sim \Gamma_0 h^4$.  Decay happens before this maximum
value is reached if
\be
\Gamma_\phi \gtrsim H_{\rm RG} \quad \Longrightarrow \quad
S_{\rm max} \gtrsim h^2
\label{dV_decay}
\ee
with $\Gamma_\phi = \Gamma_0 S_{\rm max}^2$. Eq.~(\ref{RG}) implies
large couplings, whereas Eq.~(\ref{dV_decay}) is satisfied more easily
for small couplings.  This contradiction can only be remedied for
large $n$ and small decay rates.  Decay before $V_{\rm NR}$ becomes
important does not happen for $n=4,\,5$ operators; it is marginally
consistent with $n=6$ operators in the limit $\Gamma_0 \to 10^{-26}$
and $h \to 10^{-13/2}$. The flaton mass has to be extremely small:
$m_S \lesssim \Gamma_\phi \sim H_{\rm RG} \sim 10^{-36}$.

One could contemplate the possibility that the growth of $S$ is halted
by non-renormalizable operators, and the flaton slow rolls in the
non-renormalizable potential before decay.  But this is to no avail.
In this case the left hand side of Eq.~({\ref{dV_decay}) is
$\Gamma_\phi \sim (fD)_{\rm NR}^2 S_{\rm max}^2$, with $(fD)_{\rm NR}
\propto H$ parameterizing the evolution of $S$ and $\delta S$ in the
non-renormalizable potential.  The right hand side is likewise
proportional to $H$.  Thus for $H < H_{\rm NR}$, both sides decrease
with the same rate, and it remains impossible to satisfy
Eq.~(\ref{dV_decay}).  In other words, if decay does not happen before
the non-renormalizable potential becomes important, it will certainly
not happen during the slow roll in the non-renormalizable potential if
the right amount of density perturbations are to be produced.

\subsection{MSSM flat directions}

Can the flat directions of the MSSM play the r\^ole of $S$ for a decay
rate of the form Eq.~(\ref{NR_gamma})?  We take $m_S \sim m_{3/2} \sim
10^{-16}$.  The Yukawa couplings of the MSSM vary between $10^{-6}$
and $1$, whereas the gauge couplings $h \sim 0.1$. As discussed in the
previous section, only if the inflaton decays promptly at the end of
inflation, is low scale inflation consistent with MSSM flatons. But
what are the conditions for an arbitrary scale of inflation in the range
$10 m_S < H_* < 10^{-6}$?

Since the inflaton decay rate has to be greater than the flat direction
mass there is always a gravitino problem.

\paragraph{Case 1}
Prompt decay occurs for $10^{-6} < \Gamma_0 H_* < 10^3$, see
Eq.~(\ref{P_prompt}).  This is not possible for 1FI for any scale of
inflation. For 2FI prompt decay is possible for $H_* \gtrsim 10 m_S$,
$M \lesssim 10^2 H_*^{3/4} < 1$, and $S_* < M$, which is hardest to
satisfy in the limit $H_* \to H_{\rm max} \sim 10^{-6}$.

Low scale inflation, with the Hubble constant of the order of the
gravitino mass, comes at the cost of introducing a new scale $M$
which cannot be identified with any of the known scales, such as the
SUSY breaking scale, the GUT scale, or the Planck scale.  The scale
$M$ can be identified with the scale of inflation, $M \sim \sqrt{H_*}$
for $H_* \gtrsim 10^{-8}/(\beta^2 \lambda_0^4)$.

\paragraph{Case 2}
For smaller $\Gamma_0$ the inflaton decays well after the end of
inflation, and cases 2 and 3 apply. The decay rate is greater than the
MSSM soft mass for $H_* \gtrsim 10^{-10} (fD)^{-1} \Gamma_0^{-1/2}$.
Requiring $H_* < 10^{-6}$ and decay after the end of inflation
constrains $10^{-4}H_*/(fD)^2 \lesssim \Gamma_0 \lesssim
10^{-6}/(D^2 H_*)$, where we have taken $\alpha \approx 1/6$.

Let's first consider case 2.  There are three ways to avoid thermally
induced early oscillations.  The first possibility is that the
plasmons coupling to $S$ fall out of equilibrium before plasma effects
become important, and the non-equilibrium thermal mass is small
Eq.~(\ref{RG_bnd}).  However, for MSSM gauge couplings the
non-equilibrium thermal mass is never small, and this possibility is
excluded.

The second possibility for the thermal effects to be negligible is
that the plasmons are in equilibrium until inflaton decay, and the
plasma mass is small, Eq.~(\ref{pl_bnd}).  Plasma masses are
negligible for $h \lesssim \sqrt{H_*}$, which excludes MSSM gauge
couplings.

The third possibility is that the thermal plasma is initially far from
an equilibrium distribution.  However, for gauge particles the thermal
effects can be delayed at most until $H_{\rm del} \gtrsim 0.1 H_*$.
This offers not much perspective either.

\paragraph{Case 3}
Just as for low scale inflation, all thermal constraints are stronger
for case 3 if $fD <1$.  Hence, case 3 is likewise incompatible with
MSSM flatons. The only possible exception are MSSM flat directions
whose potential is dominated by a tachyonic RG mass, i.e.,
Eq.~(\ref{Vth}) with $c_T < 0$, so that $fD >1$ is possible.  The
zero-mode grows exponentially once the tachyonic mass exceeds the
Hubble constant.  As discussed in the previous subsection, the
inhomogeneous reheating scenario can only work if the inflaton decays
before the growth is halted by the non-renormalizable potential, and
Eq.~(\ref{dV_decay}) is satisfied.

Oscillations induced by the plasma mass reduce the flaton amplitude.
As a result $H_{\rm eq} \propto H^{14/5}$, and the plasmons remain in
equilibrium until decay.  The only way out is that (some of the) the
particles coupling to $S$ are out of equilibrium from the start
$H_{\rm eq} \gtrsim H_{\rm end}$, or
\be
\Gamma_0 \lesssim 10^6 h^4 H_*.
\label{Heq_H}
\ee
The MSSM flaton couples to several fields.  If one of these couplings
$h'$ is small so that the above equation is not satisfied, then in
addition $H_{\rm RG} > H_*$ and $H_{\rm RG} > H_{\rm pl}$ for the RG
induced mass to dominate over the plasma mass from the start.  This is
only possible for flat direction with top Yukawa interactions $h \sim
1$; in all other cases Eq~(\ref{Heq_H}) should be satisfied for all
flaton couplings.

Consider first MSSM flat directions with top Yukawas.  The RG induced
potential can dominate over all other contributions to the potential
immediately after the end of inflation.  The zero mode and
fluctuations grow exponentially.  Decay happens before the
non-renormalizable potential becomes important for $H_* \lesssim
\Gamma_0 \lesssim 10^6 H_*$ and $\Gamma_0 \gtrsim 10^{-9}, \,
10^{-12}$ for $n=5, \,6$ operators; it is not possible for $n=4$
operators.

For flat directions without top Yukawa couplings the growth of the
zero-mode starts at $H_{\rm RG} \sim h^4 \Gamma_0 \lesssim H_*$.
Moreover, Eq.~(\ref{Heq_H}) should be satisfied.  These constraints
together are severe, prohibiting decay before $V_{\rm NR}$ becomes
important for $n \leq 5$, whereas it is only marginally allowed for
$n=6$ in the limit $\Gamma_0 \sim H_0 \to 10^{-6}$.


\section{Constant decay rate}
\label{S_const}

\paragraph{Renormalizable couplings}
Consider a superpotential coupling of the form~\cite{FDR}
\be 
W = \lambda \phi H_u H_d
\ee
with the coupling 
\be
\lambda = \lambda_0 \left[ 1+ \( \frac{S}{M} \)^q + ...\right]
\ee
with the ellipses denoting higher order terms.  Since the higher order
corrections play no r\^ole we will omit them in the following. In
supersymmetric and string inspired models, all couplings and masses
are, rather than being constants, functions of scalar fields in the
theory. The higher order corrections can also arise from
non-renormalizable operators, such as those in Eq.~(\ref{NR_op}). Within
the MSSM the above coupling is the only gauge invariant possibility.

The decay rate is
\be 
\Gamma_\phi = \Gamma_0 \left[ 1+ \(\frac{S}{M} \)^q \right]^2, 
\qquad \Gamma_0 = \frac{\lambda_0^2}{8\pi} {m_\phi}.
\label{R_gamma}
\ee
which is of the form Eq.~(\ref{C_gamma}) with $q \geq 1$, $r=2$ and
$M$ some cutoff scale. As before $m_\phi = \beta H_* \, (\beta
\sqrt{H_*})$ for 1FI (2FI) with $\beta \lesssim 1$. The 
decay rate is bounded from below by BBN
\be
\Gamma_0 \gtrsim {10^{-43}}
\label{R_BBN}
\ee
In addition the mass of the inflaton should be larger than that of the
decay products, otherwise decay is kinematically forbidden.  

The density perturbations are given by Eq.~(\ref{C_Sstar}).  All
scales in the problem have to lie close together, $H_* \lesssim
10^{-4}M/( \alpha f^q D)$ and $H_* < S_* < M$.  The amount of
fine-tuning is increased by inefficiencies, when $\alpha f^q D <
1/6$. Considering the density perturbations in terms of the variable
at the time of decay, Eq.~(\ref{C_S}), similarly leads to the
conclusion that the amount of fine tuning is increased also in the
opposite limit $f^q D >1$.

\paragraph{Phase space factor}
Consider 2-body decay of the inflaton, through a coupling of the form
\be 
W = \lambda_0 \phi \psi \psi
\ee
The decay rate is 
\be 
\Gamma_\phi = \Gamma_0 \sqrt{1- \( \frac{2 m_\psi}{m_\phi} \)^2},
\qquad  \Gamma_0 = \frac{\lambda_0^2 }{8 \pi} {m_\phi},
\label{PS_gamma}
\ee
where the square root comes from integration over phase space.  We
will refer to this decay rate as the phase space (PS) decay rate. If
the mass of $\psi$ is set through a coupling to a flat direction,
i.e., $m_{\psi} = \lambda S$, the decay rate is of the form
Eq.~(\ref{C_gamma}) with $q=2$, $r=1/2$ and $M = m_\psi /(2 h)$. Now
$M$ is not a fundamental scale in the problem.  The density
perturbations are given by Eq.~(\ref{C_Sstar}). Also in this case $M
\sim 10-10^3 H_*$. This gives the relation
\be
\beta \sim \left \{
\begin{array}{lll}
& 10 - 10^3 \lambda, & \quad ({\rm 1FI}), \\
& 10 - 10^3 \lambda \sqrt{H_*}, & \quad ({\rm 2FI}).
\end{array}
\right.
\label{beta_max}
\ee

The lower bound on $\Gamma_0$ comes from BBN, Eq.~(\ref{R_BBN}).
There is also an upper bound, from $\lambda_0 < 1$ and $\beta \lesssim
1 \,(10^2 \sqrt{H_*})$ for 1FI (2FI).

Note that the back reaction of $\psi$ on the flat direction, induces a
thermal mass for $S$. Said in another way, the couplings of the flaton
field to which we refer as $h$, and which control the strength of the
thermal effects, include also $\lambda$.

\subsection{Low scale inflation}  

Can the inhomogeneous reheating scenario work for low scale inflation
$H_* \sim m_{3/2} \sim 10^{-16}$ for decay rates of the form
Eqs.~(\ref{C_gamma},~\ref{R_gamma},\,~\ref{PS_gamma})? To obtain the
right density fluctuations requires $S_* \sim 1 - 10^2 H_*$ and $M
\sim 10 -10^{3}H_*$ for all examples considered above. The less
efficient the mechanism --- $\alpha (S_*/M)^q $ small --- the closer the
scales lie together. $M$ cannot be identified with inflationary scale,
and thus introduces a new scale in the problem.  This is not the case
for the PS decay rate, where $M$ is related to the effective mass of
the inflaton decay products, and does not represent a fundamental
scale.

The decay rate is bounded by the BBN constraint and the scale of
inflation $10^{-43} \lesssim \Gamma_0 \sim 10^{-18} (10^{-10})
\lambda_0^2 \beta$ for 1FI (2FI). For the PS decay rate $\beta <
\beta_{\rm max} \sim 10^{-6}$ for 2FI as follows from
Eq.~(\ref{beta_max}).  There is a gravitino problem for $\Gamma_0
\gtrsim 10^{-19}$.

\paragraph{Case 1}

Prompt decay is possible for $10^{-16} < \Gamma_0 < 10^{-13}
(H_*/M)^q$.  Small values for $q$ are favored. Note that for the upper
limit, which saturates the Gaussianity constraint, $H_* \sim S_*$ and
thus $H_*/M \lesssim 0.1$.  Prompt decay is not possible for 1FI,
whereas for 2FI it constrains $10^{-6} \lesssim \beta \lambda_0^2
\lesssim 10^{-3} (H_*/M)^q$. It is also is also incompatible with a
PS decay rate.

The window of allowed inflaton mass and couplings is much smaller than
for the polynomial decay rate.  One reason is that the perturbations
are transferred less efficiently to the radiation bath, due to the
$(S_*/M)^q$ suppression factors.  The second reason is that the
constraint $1 \lesssim \Gamma_\phi/H \lesssim 10^3 H_*$ confines
$\beta \lambda_0$ also within three decades.  This is in contrast with
the polynomial decay rate, where due to the $S$, and thus $\alpha$,
dependence of $\Gamma_\phi$, the parameter combination $\beta
\lambda_0/M^2$ is constraint only within nine decades.
 
The other flat direction couplings are unconstrained, whereas its mass
$m_S \lesssim 10^{-17}$ for a scale invariant spectrum.  The effective
mass from the non-renormalizable potential is also sufficiently small
for Planck suppressed operators with $\kappa, \, \Lambda \sim
1$. However, if these operators are generated at the same scale as the
effective cutoff, $\Lambda \sim M$, then the non-renormalizable
operators are only marginally consistent for all $n$.

The gravitino problem has to be addressed.

\paragraph{Case 2}
For smaller decay rates, $\Gamma_0 < 10^{-16}$, decay occurs well
after the end of inflation, which brings us to case 2 and 3.
Non-renormalizable operators play no r\^ole for $\Gamma_0 > 10^{-28}
\kappa/\Lambda, \; 10^{-43} \kappa/\Lambda^2,
\;10^{-56}\kappa/\Lambda^3$ for $n=4,\; 5,\; 6$. If $\kappa, \,
\Lambda \sim 1$, then only $n=4$ operators need to be considered.
Non-renormalizable operators with $\Lambda \sim M $ always dominate
before decay if $\Gamma_0 < 10^{-16}$.

Let's consider case 2 first.  The thermal constraints are severe.
They can be satisfied if the particles coupling to $S$ are out of
equilibrium, while the non-equilibrium thermal potential remains
unimportant, see Eq.~(\ref{RG_bnd}).  The non-equilibrium potential is
only small for couplings $\lambda,\,h \lesssim 10^{-7}$.  But such
small couplings are inconsistent with the plasmons being out of
equilibrium.

The second way in which thermal effects can be avoided, is that the
plasmons are in equilibrium, and the plasma mass is small, see
Eq.~(\ref{pl_bnd}).  The plasma mass is only negligible for couplings
$\lambda, \,h \lesssim \sqrt{\Gamma_0}$. The couplings scale with
$\sqrt{\Gamma_0}$, and must be especially small for 1FI.  The flaton
mass is bounded by $m_S \lesssim \Gamma_0$.  The constraints are
weakest in the limit $\Gamma_0 \to 10^{-16}$, i.e., when decay happens
shortly after the end of inflation.

Finally, the thermal effects are delayed if the initial plasma is far
from an equilibrium distribution.  The ensuing constraints are the
same as in the previous paragraph with the replacement $h \to 10^{-2}
h_{\rm pl}^2$.  The plasma couplings have to be small $h_{\rm pl}
\lesssim 10^{-3} (\Gamma_0 / 10^{-16})^{1/4}$.

\paragraph{Case 3}
The non-renormalizable potential generically plays no r\^ole.  If
either the zero temperature mass or the plasma mass becomes important
before inflaton decay, this implies fine-tuning the already tuned
values of $H_*,\,S_*,\,M$.  And without opening parameter space much,
see Eq.~(\ref{notmuch}).

The RG potential offers no better prospects, whether $c_T$ is positive
or negative.  The VEV $S$ is bounded by the scale of inflation $H_*$
and the cutoff $M$, both at the end of inflation and at the time of
inflaton decay.  To obtain the observed density perturbations, all
scales have to lie within four decades of each other at all times, and
consequently there is not much room for evolution of $S$ and $\delta
S$.

\subsection{MSSM flat direction}

Can the MSSM flat directions be responsible for the density
fluctuations, for a constant decay rate of the form
Eqs.~(\ref{C_gamma},~\ref{R_gamma},~\ref{PS_gamma})?  MSSM scalars
have soft SUSY breaking mass $m_S \sim 10^{-16}$, and at least one or
more gauge couplings with $h \sim 0.1$.  What are the conditions on
$H_*$ and $\Gamma_0$ for a successful scenario?

\paragraph{Case 1}

Just as for low scale inflation, prompt decay is only possible for a
2FI within a small window $10^2 \sqrt{H_*} \lesssim \beta \lambda_0^2
\lesssim 10^5 (H_*/M)^q$. The scale of inflation can be identified
with the cutoff scale only for $H_* \sim 10^{-6}$ and $M \sim
\sqrt{H_*} \sim 10^{-3}$.  In this case the inflaton mass and coupling
have to be large $\beta \lambda_0^2 \to 1$. Prompt decay is
incompatible with a PS decay rate for all scales of inflation.

The flaton coupling and mass are unconstrained, except for $m_S
\lesssim 0.1 H_*$ from scale invariance.  The effective mass generated
by non-renormalizable operators is likewise sufficiently small for
cutoffs $\Lambda > M$. 

The gravitino problem has to be addressed. 

\paragraph{Case 2}

There are two possibilities for thermal effects to be absent.  Either
the induced plasma mass is negligible, or if the plasmons are out of
equilibrium, the induced RG mass is sub-dominant,
Eqs.~(\ref{pl_bnd},~\ref{RG_bnd}). This is the case for $h \lesssim
\sqrt{\Gamma_0}$, respectively $h \lesssim \sqrt{S_*} \sim 10
\sqrt{H_*}$.  Since $H_* \lesssim 10^{-6}$, and $\Gamma_0 \lesssim
H_*$ (otherwise prompt decay), both options are inconsistent with
gauge couplings of order $h \sim 0.1$.  

The onset of thermal effects can be delayed, see Eq.~(\ref{delay}),
only for $H_* \to 10^{-6}$, and even then $\Gamma_\phi \gtrsim 0.1
H_*$.  Hence, this hardly opens up parameter space.

\paragraph{Case 3}

If decay happens well after the end of inflation, thermal effects are
always important.  Decay should follow quickly after the onset of
thermally induced motion of $S$ and $\delta S$.  This is because
Gaussianity and the cutoff constrain $\delta S$, $S$, and $M$ to lie
close to each other both at the end of inflation and at the time of
decay, and there is little room for evolution of $S$ and $\delta S$.
Therefore, $H_{\rm th} \lesssim 10^2 \Gamma_\phi$ or equivalently
$\Gamma_0 \gtrsim 10^{-2}H_*$.  Only a small bit of parameter space is
opened up, and only at the cost of fine tuning all parameters.


\section{Varying mass scenario}
\label{S_psi}

Suppose the field dominating the energy density has an effective mass
set by a coupling to a flat direction $m_\psi \sim \lambda S$, with
$\lambda$ a gauge/Yukawa coupling.  If the $\psi$ field couples to SM
degrees of freedom with a coupling $\lambda_0$ the decay width is
\be
\Gamma_\phi \sim \frac{\lambda_0^2}{8\pi}  \lambda S .
\ee
Decay is unsuppressed if $m_\psi$ is larger than the masses of the
decay products. The decay width is polynomial in $S$, of the form
Eq.~(\ref{P_gamma}), with
\be
\Gamma_0 = \frac{\lambda \lambda_0^2}{8\pi}, \qquad p =1.
\ee
The observed perturbations are obtained, if Eq.~(\ref{P_Sstar}) is
satisfied with $p=1$.  The decay rate decreases with $S$, and $m_S <
\Gamma_\psi$ for decay to take place at all. The thermal bath
originates from inflaton decay, and thus the thermal time scales are
as given in Eqs.~(\ref{Heq},~\ref{Hpl},~\ref{HRG}) with $\Gamma_\phi$
the $S$-independent inflaton decay rate.

There are important differences with the examples discussed up till
now, in which the inflaton was the FDED.  Now, a new field is
introduced in the theory, that is to dominate the energy density
sometime after inflaton decay.  The new field is accompanied by new
parameters, making the model less predictive.  There is no equivalent
of case 1, as the inhomogeneous reheating mechanism only works if
$\psi$ decays after it (nearly) dominates the energy density,
$\Gamma_\psi < H_{\rm dom}$. The thermal bath is produced by inflaton
decay, and in the period between $H_{\rm end} < H < \Gamma_\psi$,
thermal effects should be taken into account.  Thus, in contrast with
the inflaton as FDED, a thermal bath is already present {\it before}
$\psi$ domination.

\subsection{Constraints}
Here we discuss the conditions specific to the varying mass scenario,
namely the requirement of domination, and the thermal effects before
domination.

Consider first the case in which the $\psi$ quanta are initially in
thermal equilibrium. Further assume that the annihilation rate is
smaller than $\Gamma_{\rm ann} \lesssim m_\psi^2$.~\footnote
{For larger annihilation rates freeze-out occurs at lower temperature
$T \sim m_\psi/20$, and the number density $n_\psi$ is Boltzmann
suppressed.  Domination then happens at Hubble constants $H_{\rm D}
\ll m_\psi$ and the constraint in Eq.~(\ref{FM_constr}) becomes much
stronger}
Then the $\psi$ particles come to dominate the energy density as soon
as they become non-relativistic, at $T_{\rm dom} \sim m_\psi$ or
$H_{\rm dom} \sim m_\psi^2$. As was first noted in~\cite{FM}, the
$\psi$ plasmons induce a thermal mass for the flat direction field $S$
of the form $m_S \sim \lambda T$.  Requiring this thermal mass to be
sub-dominant at all times gives
\be 
m_{\rm th} \lesssim H_{\rm dec} < H_{\rm dom} \quad 
\Rightarrow \quad S > \mpl
\label{FM_constr}
\ee
Such large VEVs are hard to reconcile with low scale inflation $H_*
\lesssim H_{\rm max}$, and with the presence of non-renormalizable
operators.  What is more, for an initial trans-Planckian VEV the flat
direction field itself will come to dominate the energy density of the
universe while still frozen, leading to an $S$-dominated period of
inflation.

The assumptions made in the above argument is that the $\psi$-quanta
reach thermal equilibrium before domination, and that the FDED decays
while $S$ is still frozen.  If either one of these assumptions is
dropped, the negative conclusion implied by Eq.~(\ref{FM_constr}) may
be avoided.  We discuss some possibilities in turn.

A) One possibility is that the $\psi$ quanta are direct inflaton decay
products, which are out of equilibrium from the beginning 
\be
m_\psi \gtrsim T \sim \sqrt{\Gamma_\phi}
\label{FM_RG}
\ee
This requires large inflaton masses $m_\phi > m_\psi$.  The initial
distribution is non-thermal, and $n_\psi /s $ will remain constant if
in addition the annihilation rate small. This means sufficiently small
$\psi$ couplings, ruling out the possibility of identifying $\psi$
with an MSSM field.  Moreover the couplings between $\psi$ and the
MSSM sector have to be sufficiently small so that $\psi$ decays after
domination:
\be
\Gamma_\psi < H_{\rm dom}.
\ee
An RG potential of the form Eq.~(\ref{Vth}) is induced, which is
negligible only for $H_{\rm RG } < \Gamma_\psi$ or
\be
h \lesssim \sqrt{S} \( \frac{\Gamma_\psi}{\Gamma_\phi} \)^{1/4}.
\ee
Once again small couplings are needed, ruling out the possibility of
identifying $S$ with a MSSM flat direction.  Otherwise, if the RG
potential does get important and $c_T > 0$, this will lead to damping,
leading to non-Gaussianity for $D \lesssim 10^{-3}$.  For $c_T < 0$ on
the other hand, $S$ decreases until the $\psi$-quanta acquire thermal
equilibrium.  This brings us straight back to the constraint
Eq.~(\ref{FM_constr}).

To summarize, small couplings are needed for the annihilation rate and
the induced thermal mass to be small, and MSSM scalars cannot play a
r\^ole.  There is a tension with the out of equilibrium condition,
which requires a large coupling.

B) The $\psi$ field dominates the energy density immediately after the
end of inflation.  This requires the inflaton to decay (almost)
exclusively into $\psi$ quanta. Note that although
Eq.~(\ref{FM_constr}) is trivially avoided in this way, soon after the
end of inflation either the plasma mass or the RG potential generated
by the coupling $W = \lambda S \psi \psi$ will become important.  The
situation is the same as for a polynomial decay rate, as discussed in
section V, except that case 1 is excluded.

C) The inflaton decays non-perturbatively, through resonance effects.
The initial distribution of $\psi$ particles is far from equilibrium,
as well as all the other particles in the universe.  The highly
non-linear, non-equilibrium character of preheating makes it
impossible to make any definite predictions.

D) The rate of number changing interactions is small, and chemical
equilibrium is not attained.  The initial distribution is non-thermal
if the $\psi$'s are direct inflaton decay products.  The plasma mass
$m_{\rm pl} \sim h n_\psi /E_\psi$ can be smaller than its equilibrium
value.  However, to avoid that $H_{\rm dom} \propto n_\psi E_\psi$ is
likewise smaller, and Eq.~(\ref{FM_constr}) is not ameliorated, it
should be $ E_\psi$ exceeding its equilibrium value rather than
$n_\psi$ being below its equilibrium value.

E) The thermal mass $m_{\rm pl} \sim \lambda T$ does become important
before inflaton decay.  The flaton field starts oscillating in the
potential well with decreasing amplitude $S \propto H^{7/9}$, see
Eq.~(\ref{f_mth}).  If $\psi$ decays during the thermally induced
oscillations, i.e., before the zero temperature potential becomes
important, then $f \sim (\Gamma_\psi(S_*)/H_{\rm pl})^{7/2}$ and $D
\sim 1$.  Unless $\Gamma_\psi(S_*) \to H_{\rm pl}$ decay is delayed,
and the thermal constraints, as well as the constraint $m_\psi <
\Gamma_\psi$, are much stronger.

\subsection{Model building}
The varying mass scenario, in which the FDED is not the inflaton but
some other field, is more elaborate than scenarios in which the FDED
is the inflaton.  There is an extra step: the production and
subsequent domination of $\psi$ quanta.  Not only does the
introduction of an extra field lead to more parameters, and thus to
less predictability, it also introduces extra constraints.  In
particular, there is already a thermal plasma before the domination of
$\psi$, and thermal effects should be taken into account.

If the initial distribution of $\chi$ quanta is an equilibrium one,
the scenario does not work, because of Eq.~(\ref{FM_constr}).  The
natural way out is to assume the $\psi$ quanta never are in thermal
equilibrium.  However, non-equilibrium thermal effects still play a
role, and it requires small coupling and/or tuning to make it work.

The fluctuating mass scenario is not advantageous for low scale
inflation.  Prompt decay at the end of inflation is not possible, as
decay should occur only after the $\psi$-quanta come to dominate the
universe.  Late decay implies strong thermal constraints, and only
very small flaton masses and couplings are consistent.  The prospects
are much worse than for the inflaton as FDED.

Thermal constraints, which already play a role before domination, can
generically only be avoided for sufficiently small couplings.  This
excludes the possibility of identifying $\psi$ with an MSSM scalar, as
well as the possibility of identifying the flaton $S$ with an MSSM
flat direction.


\section{Conclusions}
\label{S_concl}

In the inhomogeneous reheating scenario not the inflaton but some
other field is responsible for the observed density perturbations.  It
would be economical if the new fields and scales introduced in this
scenario could be identified with the fields and scales appearing in
our models of particle physics.  In particular, we have addressed the
following two questions.  Is low scale inflation possible, with the
Hubble scale of the order of the gravitino, such that the inflaton
sector can be naturally identified with the SUSY breaking sector?  Can
any of the MSSM flat directions be responsible for the density
fluctuations?

We discussed various decay rates, obtained from non-renormalizable
couplings, renormalizable couplings, and from phase space effects. For
the last two examples, the observed density perturbations can only be
obtained if the Hubble constant, the flaton VEV, and the cutoff scale
all lie within 4 decades.  This requires some tuning, especially since
Gaussianity of the perturbations is only assured for $S_* \gtrsim 10
H_*$.  For the phase space decay rate the cutoff is not a fundamental
scale in the theory. This has the advantage that there is no need to
explain the origin of this scale.  The disadvantage is that this model
is quite constraint, and for example, prompt decay is excluded.  For
the non-renormalizable decay rate there is more freedom, since the
density perturbations do not depend on the cutoff.

After inflation, already before inflaton decay has completed, there is
a dilute plasma.  Fields coupling to the flat direction field, whether
they are in equilibrium with the thermal bath or not, will induce a
thermal mass for $S$. Such thermal corrections will lead to early
induced oscillations of the flat direction field, which are
generically fatal for the inhomogeneous reheating scenario.  The
thermal effects can be avoided if all flat direction couplings are
small, which excludes MSSM flat directions.  For a polynomial decay
rate, which can originate from non-renormalizable couplings, in
addition the flaton mass has to be much smaller than the Hubble scale
during inflation.

The thermal constraints can be trivially avoided if the inflaton
decays promptly at the end of inflation, so that there is no time for
thermal effects to act.  In this case the flat direction mass and
couplings are unconstrained, except for the requirement that $m_S
\lesssim 0.1 H_*$.  The reheat temperature is high, and there is a
potential gravitino problem.  Prompt decay however, is only consistent
with a Gaussian perturbation spectrum for a small window of inflaton
masses and couplings.  The constraints are stronger for a constant
decay rate. In particular, decay rates which inherit their $S$
dependence from phase space effects are incompatible with prompt
decay.  Large inflaton mass and couplings are needed, inconsistent
with one field models of inflation.

Finally, we discussed a variation of the inhomogeneous reheating
scenario, in which not the inflaton field but another field has a
fluctuating decay rate.  After inflation, this field first has to come
to dominate the energy density, and then decay.  The prospects for
model building are bleak.  There is no analog of prompt decay: thermal
effects are always there, even before the field comes to dominate the
energy density.  Only with small couplings can one avoid the
disastrous consequences of thermally induced masses.  This excludes
the possibility of identifying either the decaying field or the flat
direction field with an MSSM field.  Moreover, there is an extra step
in the model, making it less advantageous for low scale inflation.

\section*{Acknowledgments}
The author is supported by the European Union under the RTN contract
HPRN-CT-2000-00152 Supersymmetry in the Early Universe.


\end{document}